\documentclass[conference]{IEEEtran}
\IEEEoverridecommandlockouts
\usepackage{cite}
\usepackage{amsmath,amssymb,amsfonts}
\usepackage{algorithmic}
\usepackage{graphicx}
\usepackage{textcomp}
\usepackage{xcolor}
\usepackage[a4paper, total={184mm,239mm}]{geometry}
\def\BibTeX{{\rm B\kern-.05em{\sc i\kern-.025em b}\kern-.08em
    T\kern-.1667em\lower.7ex\hbox{E}\kern-.125emX}}

\usepackage{subfigure}
\usepackage[ruled,vlined,linesnumbered]{algorithm2e}
\usepackage{glossaries} 

\newacronym{FPGA}{FPGA}{Field Programmable Gate Array}
\newacronym{HPC}{HPC}{High-Performance Computing}
\newacronym{QPU}{QPU}{quantum processing unit}
\newacronym{NIC}{NIC}{network interface controller}
\newacronym{RFSoC}{RFSoC}{Radio Frequency System-On-Chip}
\newacronym{QRM}{QRM}{Quadrant-based Rearrangement Method}
\newacronym{SLM}{SLM}{Spatial Light Modulator }
\newacronym{PS}{PS}{Processing System}
\newacronym{PL}{PL}{Programmable Logic}

\newacronym{NISQ}{NISQ}{noisy intermediate-scale quantum}
\newacronym{LDM}{LDM}{Load Data Module}
\newacronym{QPM}{QPM}{Quadrant Processing Module}
\newacronym{OCM}{OCM}{Output Concatenation Module}
\newacronym{API}{API}{application programming interface}

\makeglossaries
\begin{document}
\title{Design of an FPGA-Based Neutral Atom Rearrangement Accelerator for Quantum Computing
\thanks{This work was funded by the German Federal Ministry of Education and Research (BMBF) under the funding program \textit{Quantum Technologies - From Basic Research to Market} under contract number 13N16087, as well as from the Munich Quantum Valley~(MQV), which is supported by the Bavarian State Government with funds from the Hightech Agenda Bayern.}
}

\author{\IEEEauthorblockN{Xiaorang Guo, Jonas Winklmann, Dirk Stober, Amr Elsharkawy, and Martin Schulz}

\IEEEauthorblockA{School of Computation, Information and Technology, Technical University of Munich, Garching, Germany \\ 
Email: \mbox{\{xiaorang.guo, jonas.winklmann, dirk.stober\}@tum.de}, \{elsharka, schulzm\}@in.tum.de}}

\newcommand{\delete}[2]{\PackageWarning{Suggested change}{#1 wants to remove "#2"}\textcolor{red!80}{#2}}
\newcommand{\add}[2]{\PackageWarning{Suggested change}{#1 wants to add "#2"}\textcolor{green!70!black}{#2}}
\newcommand{\replace}[3]{\PackageWarning{Suggested change}{#1 wants to replace "#2" by "#3"}\textcolor{red!80}{#2}\textcolor{green!70!black}{#3}}
\newcommand*{\mycomment}[3][]{%
    \PackageWarning{Unresolved comment}{by #2}
    \todo[inline, size=\small, #1]{\textbf{#2:} #3}
}

\newcommand{\AMR}[2][]{\mycomment[#1, color=orange!50]{Amr}{#2}}
\newcommand{\XG}[2][]{\mycomment[#1, color=olive!50]{Xiaorang}{#2}}
\newcommand{\DS}[2][]{\mycomment[#1, color=violet!50]{Dirk}{#2}}

\maketitle
\begin{abstract}
Neutral atoms have emerged as a promising technology for implementing quantum computers due to their scalability and long coherence times. However, the execution frequency of neutral atom quantum computers is constrained by image processing procedures, particularly the assembly of defect-free atom arrays, which is a crucial step in preparing qubits (atoms) for execution. To optimize this assembly process, we propose a novel quadrant-based rearrangement algorithm that employs a divide-and-conquer strategy and also enables the simultaneous movement of multiple atoms, even across different columns and rows. We implement the algorithm on \glspl{FPGA} to handle each quadrant independently (hardware-level optimization) while maximizing parallelization. To the best of our knowledge, this is the first hardware acceleration work for atom rearrangement, and it significantly reduces the processing time. This achievement also contributes to the ongoing efforts of tightly integrating quantum accelerators into High-Performance Computing (HPC) systems.

Tested on a Zynq RFSoC \gls{FPGA} at 250 MHz, our hardware implementation is able to complete the rearrangement process of a 30$\times$30 compact target array, derived from a 50$\times$50 initial loaded array, in approximately 1.0~${\mu}s$. Compared to a comparable CPU implementation and to state-of-the-art \gls{FPGA} work, we achieved about 54$\times$ and 300$\times$ speedups in the rearrangement analysis time, respectively. Additionally, the \gls{FPGA}-based acceleration demonstrates good scalability, allowing for seamless adaptation to varying sizes of the atom array, which makes this algorithm a promising solution for large-scale quantum systems.
\end{abstract}


\glsresetall

\section{Introduction}
Quantum computing, as a revolutionary computing approach, promises computational complexity advantages for particular problems~\cite{Satvik2023,guo2023}. Among all the different technologies, also referred to as physical modalities, that can be used to realize quantum computing, neutral atoms are particularly interesting since they can provide better scaling characteristics and longer coherence times~\cite{Mello2019}. 
\begin{figure}[t]
    \centering    \includegraphics[trim=0.6cm 0.5cm 0.4cm 0.4cm, clip, width=0.5\textwidth]{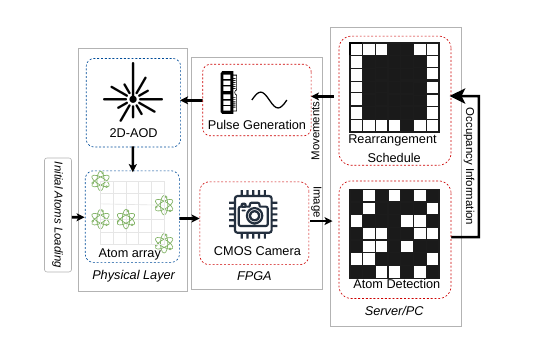}
    \caption{A typical workflow of neutral atom quantum computers. The image of the atom array is transformed into a binary representation, where black dots denote occupied areas, while white dots stand for positions where no atom is detected. This binary format serves as the input for the rearrangement algorithm. After making a schedule, the details of movements will be sent to the AWG, whose pulses control the AOD to tune the atoms.}
    \label{fig:na-workflow}
\end{figure}

In neutral atom quantum computers, preparing for quantum operations involves stochastically loading atoms into a 2D array of optical traps, with a filling probability of approximately 50\%\cite{Schlosser2002}. Then, information is commonly extracted through fluorescence imaging\cite{morgado2021quantum}, and an atom detection algorithm analyzes these images to ascertain the occupancy information of the traps. However, for reliable operation, all atoms must be arranged in a defect-free, predetermined geometric array. To achieve this fully-occupied atomic array, an atom rearrangement algorithm, or atom sorting, is essential~\cite{Wang2023}.
Starting from a random initial condition, such an algorithm develops a rearrangement schedule of the atoms to target positions, ensuring the elimination of defects in the array structure. The scheduled movements are then sent to an Arbitrary Waveform Generator (AWG) to control the Acousto-Optic Deflector (AOD) to manipulate the qubits. Fig.~\ref{fig:na-workflow} summarizes the workflow of the neutral atom quantum computer. 

\begin{figure}[bt]
    \centering
    \includegraphics[width=0.50\textwidth]{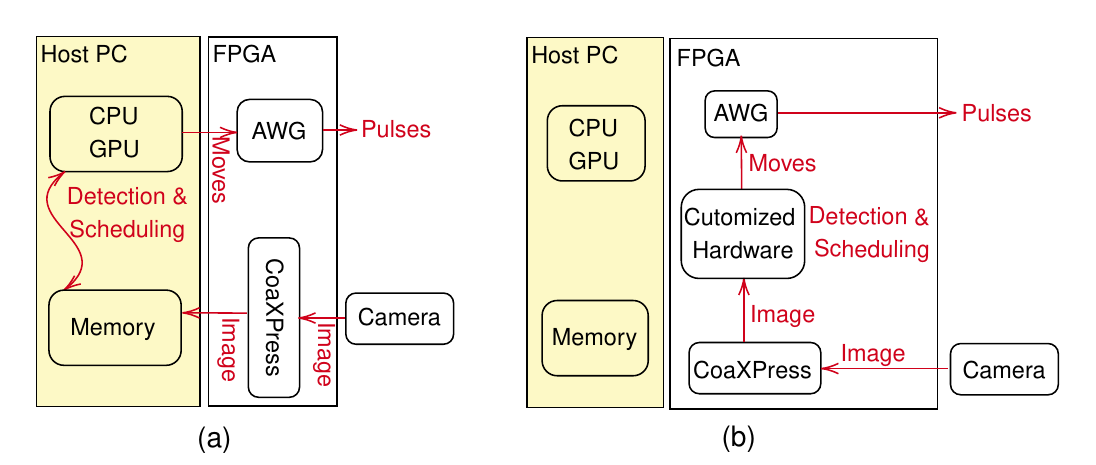}
    \caption{Architectures of atom control systems. (a) Current typical system structure: Detection and rearrangement processes are performed using CPUs or GPUs, where communications between components are needed. (b) Optimal (long-term) structure: All functional blocks are seamlessly integrated into \glspl{FPGA}, with detection and rearrangement implemented via customized hardware.}
    \label{fig:control-str}
\end{figure}

Although neutral atoms benefit from their long coherence times~\cite{wu2021concise}, the runtime for atom rearrangement in scaled-up systems with mid-circuit measurements remains a challenge~\cite{Wang2023}. To date, significant efforts have been directed towards the development of faster rearrangement algorithms utilizing multi-mobile tweezers to maximize simultaneous atom moves, achieving a higher level of parallelization~\cite{Wang2023,ebadi2021quantum,Tian2023}. However, one of the major challenges to enabling scalable and efficient neutral atom control systems is mitigating the communication overhead among system components. Fig.~\ref{fig:control-str}(a) demonstrates a control system architecture where camera connection via CoaXpress protocol~\cite{CoaXPress} and signal generation using AWG are performed on \glspl{FPGA}, but atom detection and rearrangement analysis are processed on a CPU/GPU located in a server. This cumbersome transmission process leads to significant overheads. To overcome this limitation, we propose a fully \textit{\gls{FPGA}-based control system} as shown in Fig.~\ref{fig:control-str}(b), which includes our focused rearrangement module.

In addition to reduced overhead, \gls{FPGA}-based rearrangement offers significant acceleration due to the inherent parallelism of the process, where the atom array's four quadrants follow a uniform rearrangement schedule. Building on this, we propose a \gls{QRM} to further accelerate the process on the \gls{FPGA}'s \gls{PL} side. This approach not only speeds up the rearrangement algorithm, but also reduces communication overhead with other system components.

This paper mainly focuses on the \gls{QRM} algorithm and its \gls{FPGA} implementation, and has the following contributions:
\begin{itemize}
    \item We identify a uniform moving pattern within an atom array during the rearrangement process and introduce a quadrant-based acceleration-friendly algorithm.
    \item We develop, to the best of our knowledge, the first work to implement the rearrangement algorithm on the \gls{PL} part of an \gls{FPGA}, achieving high parallelization and very low processing latency.
    \item We show that our \gls{FPGA} accelerator achieves about 54$\times$ speedup in terms of processing latency compared to a CPU implementation when tested using a 50$\times$50 atom array. Furthermore, compared to the recent \gls{FPGA} rearrangement design~\cite{Wang2023} (only implemented on the ARM core on the \gls{FPGA}), we achieve a speedup of around 300$\times$.
\end{itemize}

\section{Background} \label{sec:Background}


\subsection{Neural Atom Basics}
Neutral atoms trapped in optical lattices present a promising platform for implementing quantum computers. In this approach, qubits are encoded in individually trapped atoms, with interactions mediated via their electronically highly excited Rydberg states~\cite{Jaksch2000,morgado2021quantum}. By applying crossed laser beams to the qubits, we induce periodic transitions between the two quantum states, '0' and '1', resulting in gradual changes in the probability of the atom's state. Consequently, the atom can enter any superposition state at specific time points. Entanglement between two qubits can be achieved through the Rydberg blockade mechanism~\cite{wu2021concise}. By simultaneously manipulating two atoms and exciting their electrons to a highly energetic state, the influence between them prevents exactly one of them from entering the excited state, thus inducing entanglement~\cite{graham2019rydberg}. 

For quantum computing purposes, we need to generate a regularly spaced atom array that can be prepared by a 2D \gls{SLM}~\cite{Barredo2016}. SLMs are able to guide a cloud of cold atoms into targeted traps to load atoms. However, this process is probabilistic, with a loading probability (each optical trap in the array can successfully capture and hold a single atom during the loading process) of around 50\%. Therefore, we need a rearrangement process to construct a defect-free array for execution.

\subsection{Hardware Constraints}
To control the movements of atoms, we provide our 2D-AOD setup with several RF frequencies per dimension, thereby allowing us to generate a grid of movable tweezers. This kind of multi-tweezers system is able to move atoms at the same time when they are to be moved towards the same direction with the same step size. While this enables us to parallelize the atom rearrangement, it greatly complicates the process. Since the incident beam passes through two subsequent AODs, we can only choose a set of rows and columns instead of individual traps. A trap will be generated at every coordinate for which both the row and column are selected. If we want to generate a trap at location $(x_1,y_1)$ and another at $(x_2,y_2)$ by selecting $x_1$, $x_2$, $y_1$, and $y_2$, then there will also be traps at $(x_1,y_2)$ and $(x_2,y_1)$. In situations where this constitutes a problem, the two atom sites will have to be addressed in separate moves. Having generated such a set of traps, we can shift the contained atoms in lockstep to a target location.

\section{Methodology}
\label{sec:Algorithm}
To reduce the processing time of the atom rearrangement from the algorithm perspective, we first analyze the typical rearrangement procedure and identify the inherent parallel moving pattern. Then, we propose an efficient and acceleration-friendly algorithm to reduce the time frame for schedule analysis. 
\subsection{Typical Rearrangement Procedure} 
\begin{figure*}[tb]
    \centering
\includegraphics[width=1.0\textwidth]{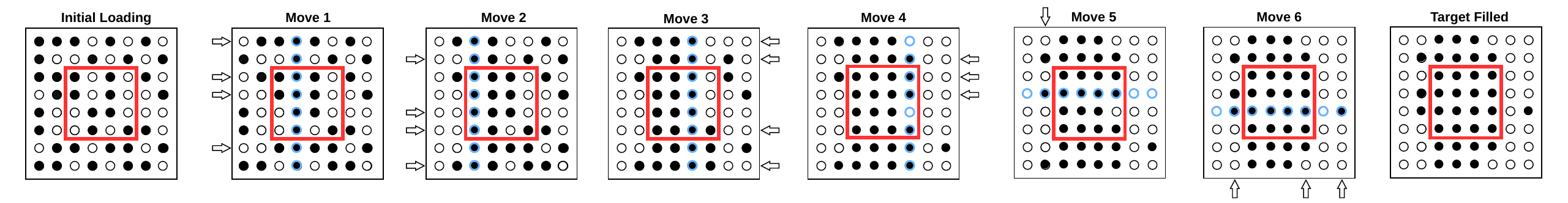}
    \caption{Illustration of the typical rearrangement algorithm. The blue emphasized line indicates the target line for this move step. Arrows represent the atoms being moved in this step, with the direction of arrows indicating the direction of movement. The red square stands for the target filling area. Within one "Move" block, we can have multiple simultaneous moves. For instance, in "Move 1", there are empty holes in Rows 1, 3, 4, and 7, so we move all atoms positioned to the left of each hole, shifting them one step to the right.}
    \label{fig:rearrangement}
    \vspace{-2mm}
\end{figure*}

\begin{figure*}[tb]
    \centering    \includegraphics[width=1.0\textwidth]{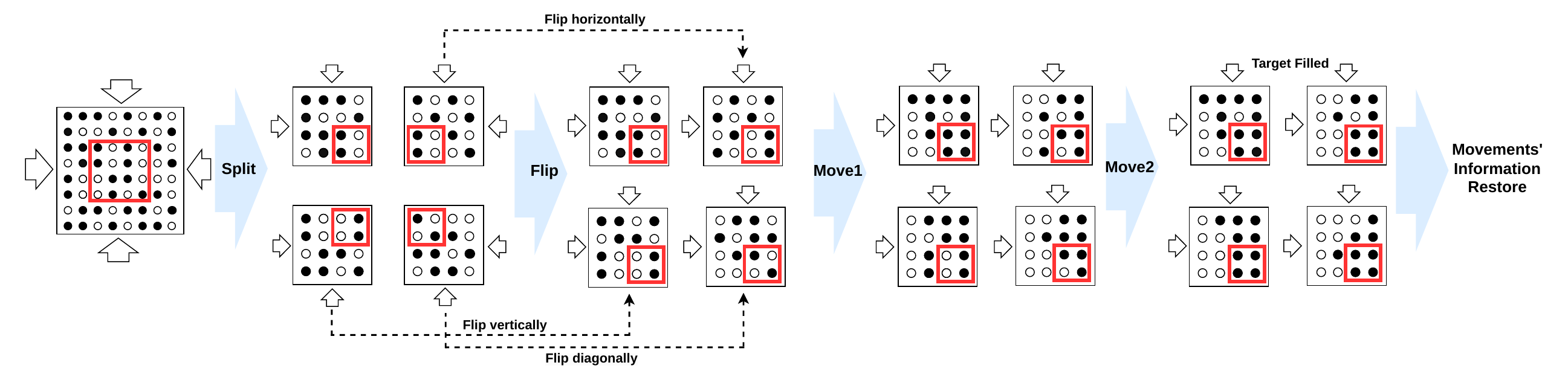}
    \caption{\gls{QRM} rearrangement schedule. By splitting the atom array and performing specific flip operations, we can apply a unified rearrangement method to each quadrant. This method provides an inherent acceleration on \gls{FPGA}.}
    \label{fig:QRM}
\end{figure*}

Atoms are loaded stochastically into a large array, and the target area is typically located in the center (due to physical fidelity constraints). The key point of the rearrangement procedure is to reposition (compress) atoms to the array's center. Fig.~\ref{fig:rearrangement} illustrates the compression mechanism of a typical algorithm using an 8 $\times$ 8 square lattice as a demonstration. Initially, we construct a matrix to represent the lattice, where each cell can either be occupied by an atom (indicated by a black dot) or remain empty (indicated by a white dot). When loading the atoms, we apply a filling factor of 50\%, aiming for a defect-free array size of 4 $\times$ 4. 

The rearrangement process involves two types of movements: horizontal moves (in Fig.~\ref{fig:rearrangement}, Move 1 to Move 4) and vertical moves (Move 5 to Move 6). In the horizontal movement procedure, we begin by identifying the center column of the array. Next, we locate the empty sites within this central column. Subsequently, we shift the atoms from the left or right (depending on whether the column we select is in the right or left part) towards these empty traps, effectively filling the gaps. Notably, with the multi-tweezer hardware employed in our system, we are able to move multiple atoms simultaneously, only if they are all displaced in the same direction with the same step size. This protocol is also illustrated in Fig.~\ref{fig:rearrangement}. In "Move 1", a total of nine atoms, arranged in four distinct lines (indicated by the white arrows on the left), are shifted toward the blue lines. Once the atoms have been adjusted within the center column, we extend the same process to the adjacent columns, ensuring a consistent filling procedure. This sequence of actions continues until we reach the boundary of the targeted filling area. For the vertical moves, we utilize the same schedule strategy, but move rows in this case. This process results in the seamless filling of the holes and a parallel movement, generating defect-free columns throughout the process.

\subsection{Quadrant-based Rearrangement Method (\gls{QRM})} 
\label{sec:ssmoving}
By analyzing the typical moving procedure, we have observed a consistent atom movement pattern across the four quadrants in this process, suggesting a uniform mechanism. Compressing all the atoms to the center of the large array is essentially compressing the atoms into the corner of each quadrant. Therefore, when we flip array quadrants via a specified direction, we can reuse the same moving schedule on each quadrant, which introduces an intrinsic parallelization with a factor of four. Thus, we propose a quadrant-based rearrangement method, as shown in Fig~\ref{fig:QRM}. In \gls{QRM}, we first split the large atom array into four parts and flip each part in different directions to put the target area at the bottom left corner. Then, a unified moving schedule is applied to the four parts. After the target area is filled, we restore the moving information to its original position.

\gls{QRM} suggests a promising parallelization architecture for an \gls{FPGA} implementation: we can create four pathways, each aligned to a quadrant. By independently processing the schedule analysis within each quadrant with a single functional call, we can realize an inherent acceleration.

\section{\gls{FPGA} Implementation} \label{sec:Implementation}
Having discussed the moving schedule, we detail the implementation of the \gls{QRM} rearrangement algorithm on a Zynq \gls{RFSoC} \gls{FPGA} board\cite{rfsoc} for acceleration in this section. 

\subsection{System Overview of \gls{QRM} Accelerator}
The system architecture used for implementing our \gls{QRM} algorithm is an \gls{FPGA} board consisting of three main components: the \gls{PS} component, the \gls{PL} component, and DDR memory.
The \gls{PS} part hosts an ARM core equipped with a Linux system responsible for handling memory access and interconnect logic, which we use to control I/O and to orchestrate the rearrangement, while the 
\gls{PL} part of the board is used to hold our customized rearrangement accelerator (described in Sections~\ref{sec:sort_core} and~\ref{sec:kernelshift}). 
The DDR memory is used for communication between \gls{PS} and \gls{PL} and holds both data used and produced by the rearrangement procedure.
To enhance data transmission efficiency, we pack 1024-bit data into one packet to move the data from DDR memory into our accelerator with minimal transmission overhead.

In this work, the whole processing procedure is controlled via a Python \gls{API}, which we implement to run on the ARM part of our target system. After the user loads the initial data into the memory, they can trigger the rearrangement step. For this, we then configure an AXI-based transmission to transfer the binary data, containing the output of the atom detection unit in the form of a bitfield, via the internal bus into our rearrangement module. Inside the module, we have a buffer to temporarily store the data. Once the transmission is finished, we kick off the rearrangement algorithm on the \gls{PL} of the \gls{FPGA}, which then stores the needed atom movements back into the memory. When the rearrangement process is finished, the final matrix and movement details are transferred back and analyzed on the \gls{PS} side.

\subsection{Rearrangement Module} \label{sec:sort_core}

\begin{figure}[tb]
    \centering \includegraphics[trim=1.5cm 0.1cm 0.3cm 0.1cm, clip, width=1.1\linewidth]{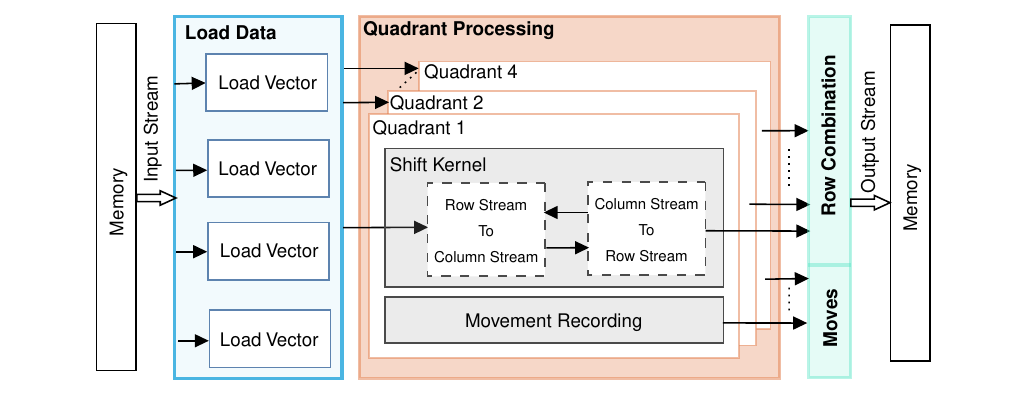}
    \caption{Complete dataflow of the HLS-accelerated rearrangement module.}
    \label{fig:hls_flow}
    \vspace{-4mm}
\end{figure}

The proposed rearrangement module for the \gls{QRM} algorithm, illustrated in Fig.~\ref{fig:hls_flow}, features inherent parallelization within the four quadrants to optimize performance. Designed in a data-flow fashion, the accelerator is fully pipelined and comprises three main modules: the \gls{LDM}, the \gls{QPM}, and the \gls{OCM}.

Initially, the input data stream is processed by the \gls{LDM}, where four \textit{Load Vector} units divide the large atom array into smaller arrays, each representing a quadrant. During this stage, the flip operation is automatically performed to prepare the data. These smaller arrays are then passed to the \gls{QPM}, enabling independent and parallel processing of each quadrant.
Within each quadrant, the core operation, referred to as the \textit{Shift Kernel}, determines the schedule for atom movements. This process will be explained in detail in Section~\ref{sec:kernelshift}. Accompanying the \textit{Shift Kernel} is a movement recording unit responsible for tracking atom movements and restoring their original positions from the quadrant-based locations. This unit records essential details, such as the original location of atoms, their directional shifts, and the number of steps taken.
Once the processing steps are completed, the movement records and the final defect-free array are consolidated into a single output stream within the \gls{OCM}. This integrated output stream is then transmitted to memory for further use.

\subsection{Shift Kernel}
\label{sec:kernelshift}

\begin{figure*}
  \centering
  \subfigure[Shift Unit after 3 cycles]{\includegraphics[width=0.38\linewidth]{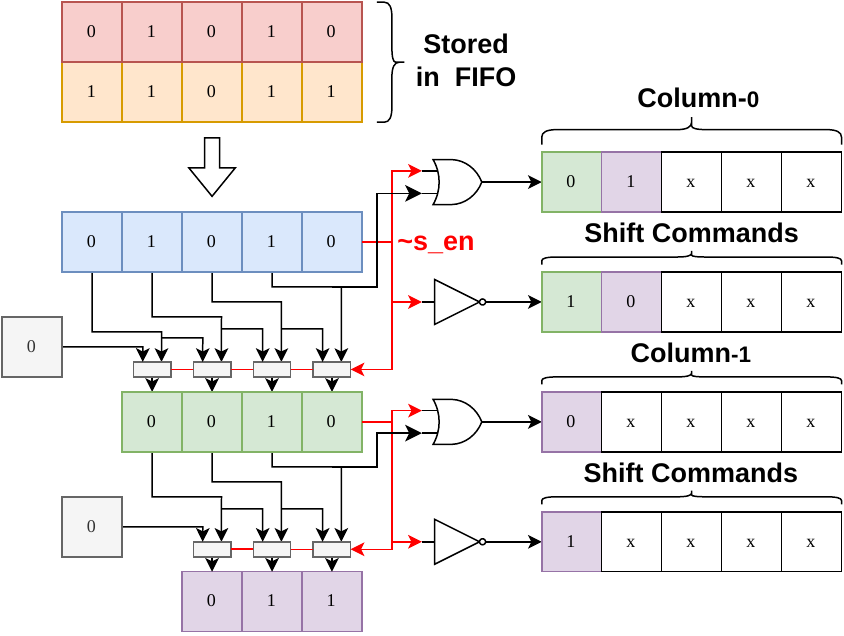}}
  \hspace{2cm}
  \subfigure[Shift Unit after $Q_w + 1$ cycles]{\includegraphics[width=0.43\linewidth,]{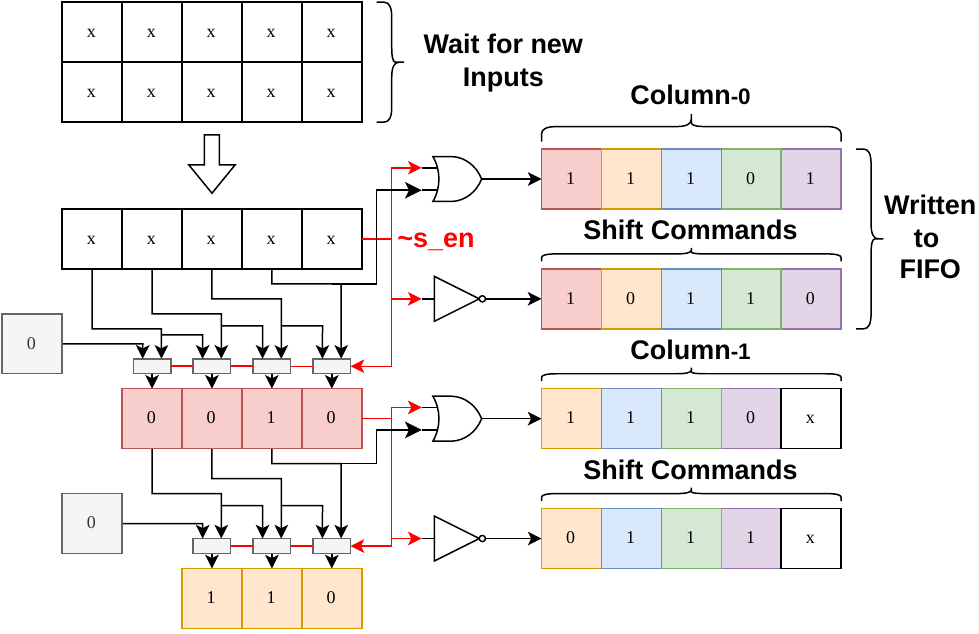}}
  \caption{Row-wise to column-wise shift process for a $10 \times 10$ input atom array ($Q_w = 5$). (a) State of the shift unit after 3 cycles. The shift kernel is processing the first three rows, and each bit in the row buffer is transferred to the corresponding \textit{column buffer}. The \textit{shift commands} buffer records whether the shifted value is '0'. (b) (b) Status after $Q_w+1$ cycles. At this time point, the purple, green, and blue rows from (a) have finished processing. \textit{Column-0} ('11101') represents the original right-most bit in each row, demonstrating the row-to-column transition process.}
  \label{fig:kernel_shift}
  \vspace{-2mm}
\end{figure*}

When migrating the rearrangement process onto \gls{FPGA}, one of the main challenges is \textit{how to efficiently perform the movements and make the movement schedule}? To solve this problem, we employ shift registers on \gls{FPGA} to represent the atom movements. By interpreting columns as rows, we design a fully pipelined shift kernel that can process both vertical and horizontal movements with minimum latency. Since different quadrants have already followed the specific flip operation in the \gls{LDM}, we can reuse the same shift architecture for all quadrants.

In this unit, every row/column is represented as a bit vector. Each quadrant has $Q_w = \frac{W}{2}$ rows, where W represents the length (size) of the initial array. The architecture can inspect every element of every row in a fully pipelined way; thus, we are able to start processing a new row every clock cycle. This results in a total latency of $Q_w$ when starting to process all rows. Inside a single quadrant, we first shift row-wise and then column-wise and repeat both steps until the center is filled. In this case, for one iteration round, the total latency of the shift kernel is $2 \times Q_w$ plus the processing time of a single row.

Fig.~\ref{fig:kernel_shift} explains the working principle of the shift kernel in detail with a row-wise example of the northwest (NW) quadrant. 
In this example, we use an atom array with a size of 10 $\times$ 10, so the sub-array for each quadrant is 5 $\times$ 5. Each row (represented by a specific color) is first fetched from the input queue for analysis, and then every element of each row is inspected step-by-step. As also shown in the figure, different rows can be analyzed simultaneously by benefiting from the pipeline mechanism to maximize the throughput and parallelization of our design.
We use the \textit{shift commands} buffer to record which atoms are shifted and the \textit{column buffer} to record the states of the atoms after the shifts. 
After fetching one row, we check its lowest significant bit (LSB). If it is set, no shift command has to be issued, and we store a '1' (the value of the LSB) into the respective \textit{column buffer} and a '0' (no shift) into the \textit{shift commands} buffer.
After processing the LSB, we shift the entire row by one to the right to check the next bit in the next stage, and we can read a new row for the current stage. Furthermore, to be able to prevent unnecessary shifts far from the center, we introduce a manual-control mechanism. In this mechanism, we can manually set the $s_{en}$ signal (shown in  Fig.~\ref{fig:kernel_shift}) to '0' to block the row elements from shifting.

After processing each quadrant, 
the four \textit{shift command buffers} are written into a large FIFO and processed by the \textit{Row Combination Unit}.
All four command buffers are processed at the same time, and it is also statically known which shift commands finish at which time.
The logic of merging the shift commands follows the rules explained earlier.
First, all '1's inside the shift command vector of a single quadrant can be executed simultaneously. For example, in Fig.~\ref{fig:kernel_shift}(b), the red, blue, and green elements in Column 0, which all have a '1' in the \textit{shift commands}, can be shifted at the same time.
Secondly, we analyze the shifts of the NW and southwest (SW) quadrants at the same time and perform the initial shifts in the same command since they contain the same shifts for the most central column from the west. 
Similarly, the shifts of the quadrants northeast (NE) and southeast (SE) can be merged as well to shift once from the east.
The column-wise \textit{shift commands} represent shifts from the south and north. Thus, the quadrants NE and NW are merged, as are SE and SW. During this process, empty shifts are removed from the final schedule to reduce the total number of commands.


\section{Evaluation}
\label{sec:Result}
\subsection{Experimental Setup}
In this work, our rearrangement accelerator is programmed on the \gls{FPGA} using an HLS-compatible C++. The design is synthesized and generated with Xilinx Vitis HLS 2022.2, and we use Xilinx Vivado 2022.2 to implement and deploy the entire project on the \gls{RFSoC} \gls{FPGA} at a clock frequency of 250 MHz. Additionally, we conduct a comparative analysis by running the C++ algorithm on the CPU. The CPU is an 11th-generation Intel Core i7 operating at 2.8GHz. As input data, we use a randomly generated matrix representing a random distribution of atoms (instead of using actual detection data from image analysis), as the randomized images have the same distribution on average and, hence, are sufficient for the analysis of our hardware design. 

\subsection{Execution Time}
\begin{figure}[tb]
  \centering
  \subfigure[Time comparison between CPU and \gls{FPGA}.]{\includegraphics[width=0.75\linewidth]{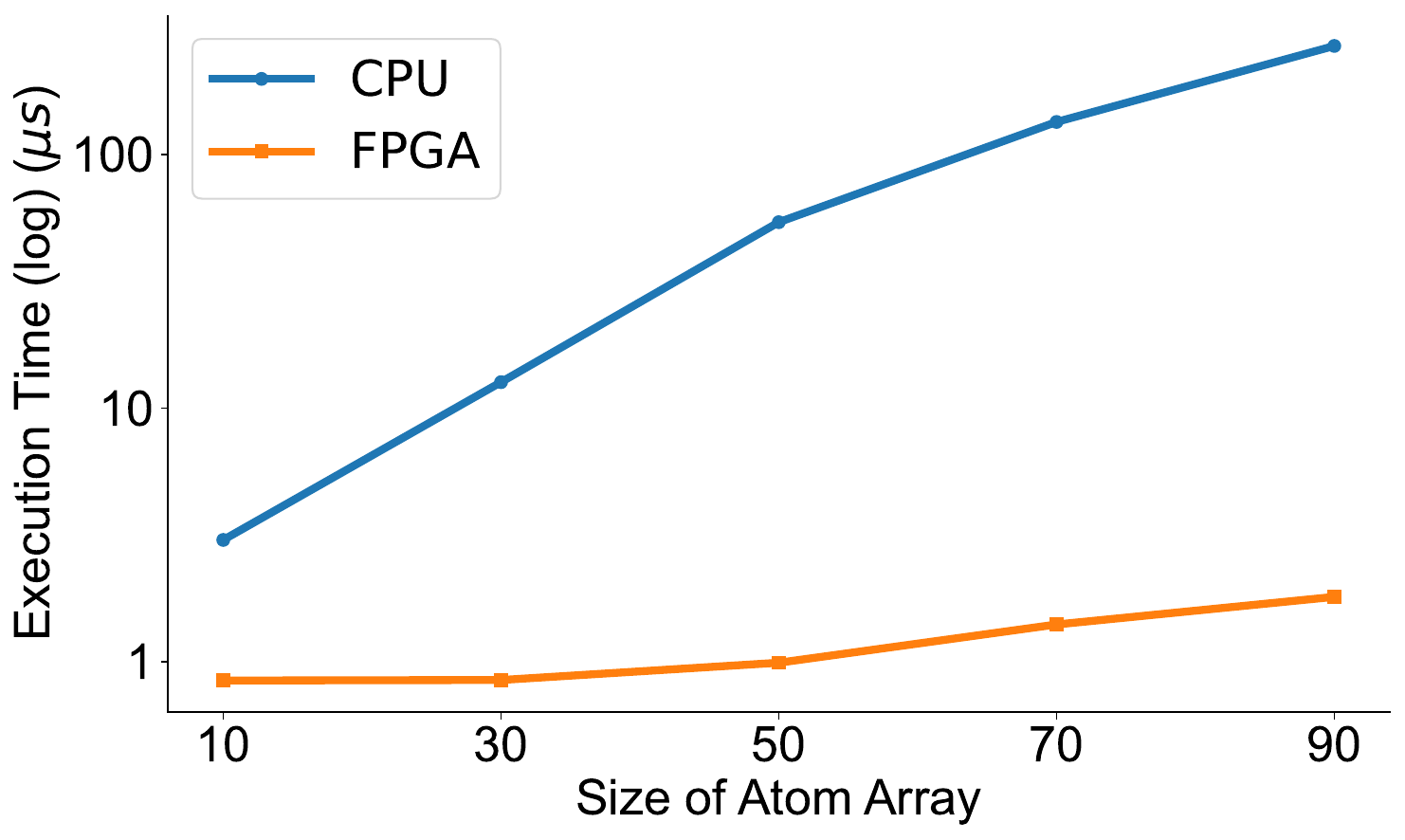} \label{fig:time_comp}}
  \quad
  \subfigure[Time benchmarks with the array of 20 $\times$ 20.]{\includegraphics[width=0.75\linewidth]{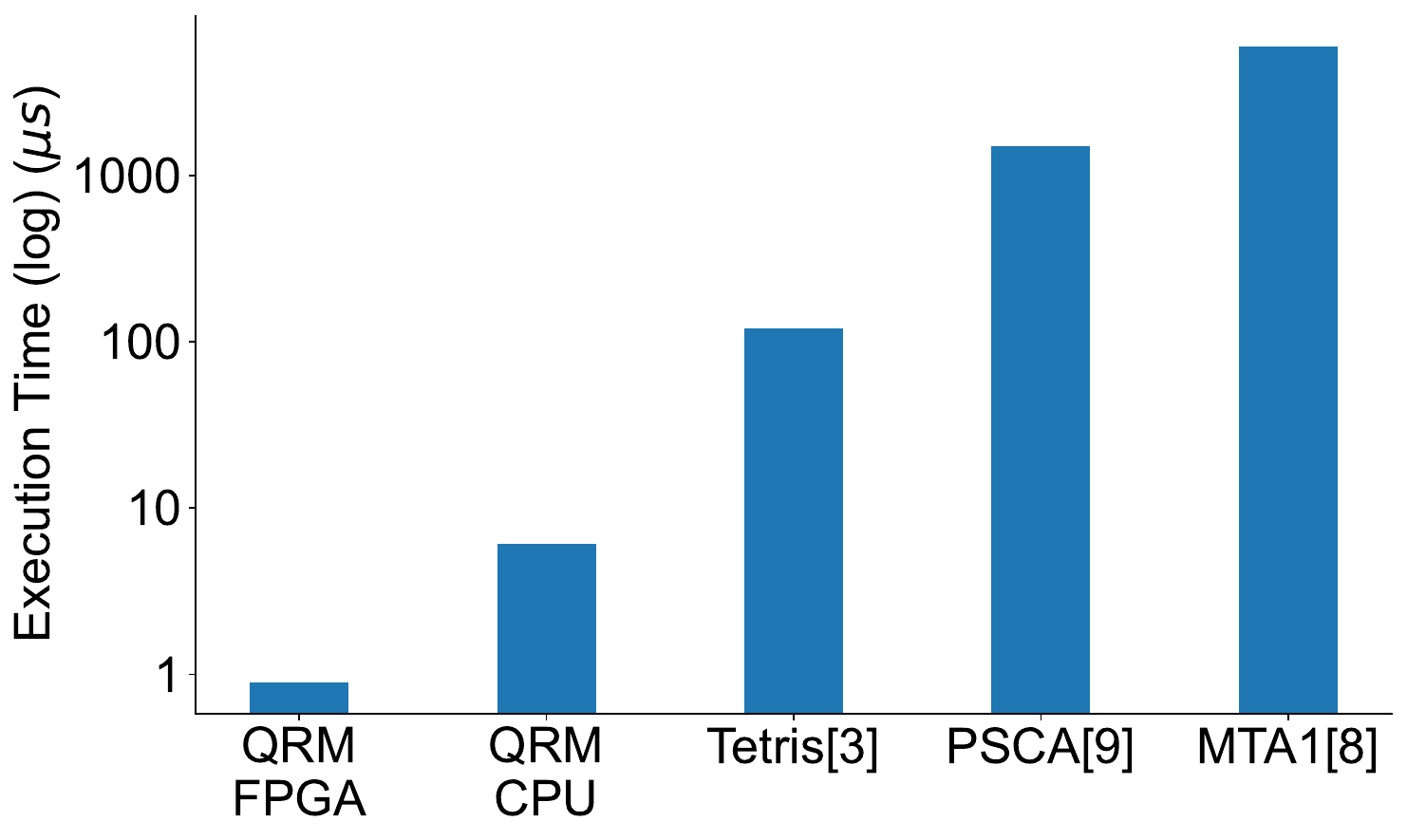} \label{fig:time_bench}}
  
  \caption{(a) Comparison of \gls{QRM} execution time between CPU and \gls{FPGA}, when scaling the size of atom array from 10 to 90 with a step size of 20. (b) Execution time comparison of different rearrangement algorithms with the initial array of 20 $\times$ 20.  The y-axis for both (a) and (b) is displayed on a logarithmic scale to accommodate the wide range of execution times.}
  \label{fig:evaluation}
  \vspace{-4mm}
\end{figure}


Fig.~\ref{fig:time_comp} illustrates a performance comparison between the CPU-based \gls{QRM} and its \gls{FPGA}-accelerated version when varying the initial atom array sizes from 10$\times$10 to 90$\times$90 with a step size of 20. As demonstrated in Fig.~\ref{fig:time_comp}, employing our \gls{FPGA}-based solution yields a substantial speedup, achieving approximately 54$\times$ faster processing when the array size is 50, and up to 134$\times$ acceleration for an array size of 90. Moreover, a noteworthy aspect of our design is its scalability, which is evident in the comparison of performance trends with the CPU version. Our design exhibits a more moderate increase in run time as the size of the initial array grows (0.8~${\mu}s$ with a size of 10, 1.0~${\mu}s$ with a size of 50, and 1.9~${\mu}s$ with size of 90), indicating its suitability for large-scale quantum computation scenarios. It's also important to note that the latency of our design is not directly dependent on the target area we aim to fill. Instead, it correlates solely with the initial size of the array and the number of iterations required by the algorithm, which is only indirectly influenced by the target size. Since we check and move atoms round by round towards the center, several iterations are necessary to completely fill the target area. Consequently, the number of movements needed increases with the target size. In our experiment, four iterations were used to complete the entire process, ensuring successful rearrangement.

We also compare the execution time of our design with other rearrangement approaches~\cite{Wang2023, ebadi2021quantum, Tian2023}, as shown in Fig.\ref{fig:time_bench}. In this benchmark, a fixed array size of 20 × 20 is used, following the experiment setup of works\cite{ebadi2021quantum, Tian2023}. The Tetris algorithm~\cite{Wang2023} is executed on an ARM core on an \gls{FPGA} platform, while the other approaches are all CPU-based. As shown in the figure, our implementations (both on CPU and \gls{FPGA}) demonstrate significant acceleration over the existing designs. More specifically, benefiting from advanced algorithm design, our CPU-based \gls{QRM} achieves an acceleration of around 20$\times$ over Tetris, 246$\times$ over PSCA\cite{Tian2023}, and almost 1000$\times$ over MTA1\cite{ebadi2021quantum}. After being implemented on the \gls{FPGA}, the performance is further improved. Compared with the best candidate (Tetris) in related works, our \gls{FPGA}-based \gls{QRM} achieves an acceleration of 120$\times$ with only 0.9$\mu s$. Furthermore, attributed to our good scalability, for a larger testing array with 50$\times$50, the acceleration can go up to 300$\times$.

\begin{figure}[tb]
    \centering
    \includegraphics[width=0.75\linewidth]{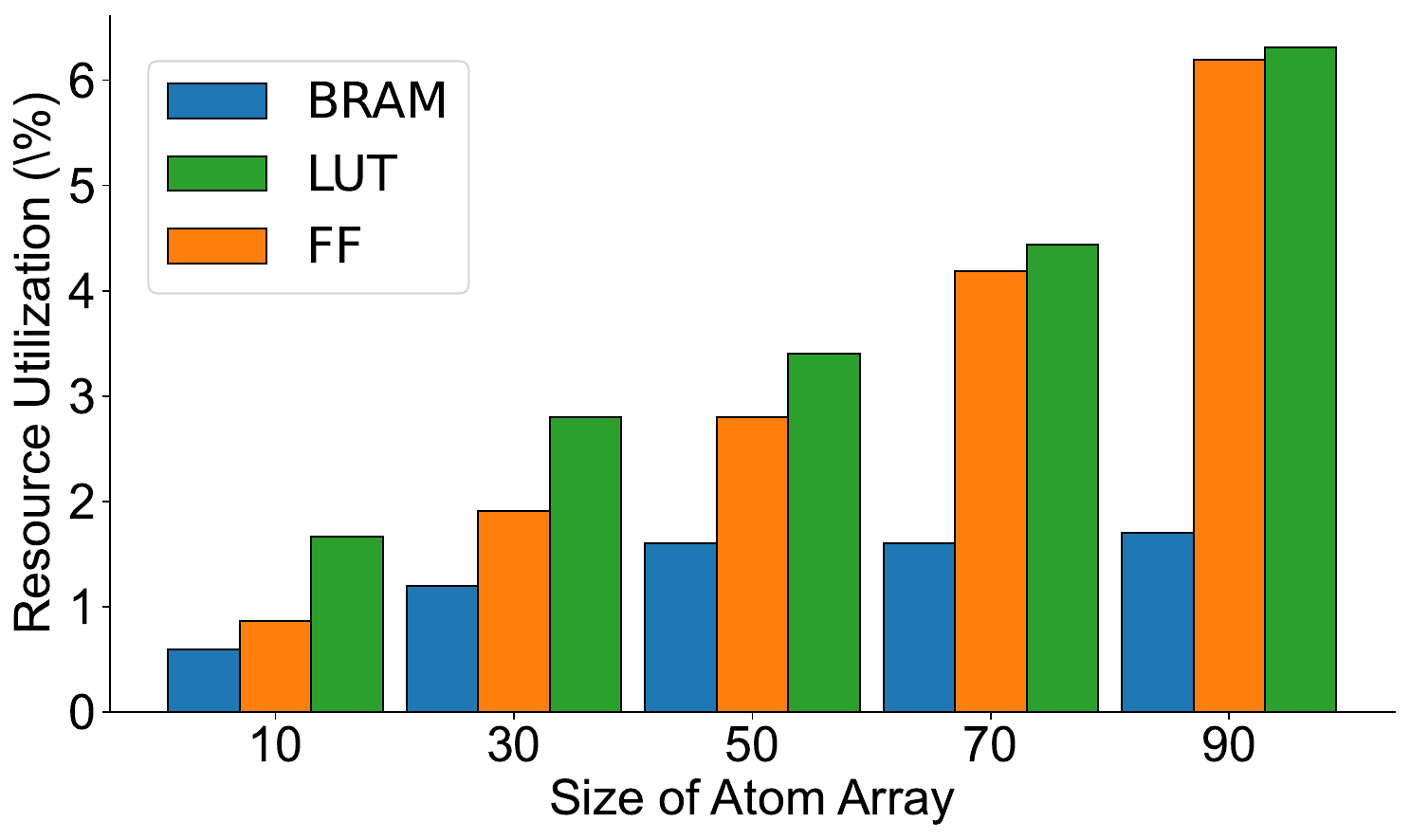}
    \caption{\gls{FPGA} resource utilization (percentage) when varying the size of atom arrays.}
    \label{fig:resource_comp}
    \vspace{-4mm}
\end{figure}

\subsection{Hardware Utilization}
Hardware resource utilization is an important metric to evaluate our design's scalability on the \gls{FPGA}. Fig.~\ref{fig:resource_comp} depicts the variation in resource utilization (percentage) with the same test scenario depicted in Fig.~\ref{fig:time_comp}. BRAM exhibits consistent behavior across different array sizes, maintaining stable consumption levels regardless of the array size ranging from 30 to 90. Both LUT and FF utilization demonstrate a linear increasing trend, with FF increasing slightly faster than LUT. However, even with a 90$\times$90 initial array, the LUT consumption remains at 6.31\%, and FF consumption at 6.19\%, ensuring enough space for other essential functional blocks required for neutral atom quantum computation or even other control logics for other physical modalities (cross-technology control scheme). 
More specifically, only about half of the resources are occupied by the four \gls{QPM}, and the other half belongs to the logic to integrate the outputs, which can be optimized in future work. 

In summary, the collective analysis of the results presented in Fig.~\ref{fig:time_comp} and Fig.~\ref{fig:resource_comp} illustrates the great scalability of our design. It exhibits high adaptability in terms of both latency and hardware performance, effectively accommodating variations in the atom array size.

\section{Conclusion}
\label{sec:Conclusion}
In this work, we presented an \gls{FPGA}-based atom rearrangement algorithm named \gls{QRM}. The main effort of this work focused on an efficient moving schedule development on the algorithm side and an efficient architecture design on the hardware side. This hardware-software codesign approach significantly reduces the analysis time of atom rearrangement and guarantees a lower clock cycle of neutral atom quantum computers. 
We evaluated our work with the input atom array of 50 $\times$ 50 on a Zynq \gls{RFSoC} \gls{FPGA}, and our implementation can finish the rearrangement schedule analysis with around 1.0~${\mu}s$. The result indicates a 54$\times$ improvement compared to a CPU implementation, and 300$\times$ speedup compared to state-of-the-art \gls{FPGA} work (Tetris algorithm). Furthermore, we tested the execution time and hardware consumption across different sizes of input atom arrays, revealing great scalability, low analysis latency, and efficient hardware utilization.

\bibliographystyle{IEEEtran}
\bibliography{ref}
\end{document}